\documentclass{ws-p8-50x6-00}

\begin{document}

\title{Caustic rings and cold dark matter}


\author{Ben Moore}

\address{Department of Physics, Durham University, Durham, DH1 3LE, UK
\\E-mail: ben.moore@durham.ac.uk\\www.nbody.net}

\maketitle

\abstracts{
The hierarchical cold dark matter (CDM) model for structure formation
is a well defined and testable model.  Direct detection is the best
technique for confirming the model yet predictions for the energy and
density distribution of particles on earth remain inadequate.  Axially
symmetric collapse of collisionless dark matter can leave observable
caustic rings in phase space and this model is frequently used to make
experimental predictions (Sikivie 1999).  Such cold collapses
inevitably suffer from radial orbit instabilities that produce
unrealistic bar-like halos.  Moreover, this model bears no relation to
the hierarchical growth of CDM galactic halos which form
via a complicated sequence 
of mergers and violent relaxation. This process destroys any symmetry and 
phase wraps existing caustics on a scale comparable to  
the first objects to collapse.  Since axions can cluster on microscopic
scales and free streaming of neutralinos only erases structure smaller
than $\sim 100 (GeV/m_{_{CDM}})$ A.U., the dynamical
effects of caustics in the Galactic halo are expected to be 
negligible.
}

\section{Introduction}

Revealing the nature of dark matter is fundamental to cosmology and
particle physics. A combination of observations and theory suggests
that the dark matter consists of non-baryonic particles, and in this
large class of hierarchical cosmogonic models a universe dominated by
cold dark matter (CDM) remains plausible albeit with some potential
problems on small scales.

The only convincing method for confirming the existence of
non-baryonic dark matter is by direct detection. Many such experiments
are in progress and are beginning to probe the parameter space allowed
by cosmological and particle physics constraints (e.g. DAMA, CDMS).
Direct and indirect detection experiments are highly sensitive to
the local density of particles and their velocity distribution. For
example, the flux of gamma-rays on Earth from neutralino annihilations
in the galactic halo depends on the amount of substructure in the dark
matter (e.g. Bergstrom etal 1999). 
It is therefore crucial to understand the phase space
structure of galactic halos in hierarchical models in order that
experiments can be fine tuned to search for the appropriate signals,
and that event rates or modulation effects can be correctly interpreted 
(Ullio etal 2000).

Many of the ongoing dark matter searches adopt the principle
that CDM particles passing through the solar system have a smooth
continuous density distribution with isotropic Maxwellian velocities.
Other halos models have been studied e.g.  Sikivie (1992, 1999),
who assume axially symmetric and cold collapse of matter to infer the
presence of caustic rings in the solar neighbourhood. We can now use
the results of high resolution computer simulations of structure
formation within a CDM universe to examine these assumptions more
carefully.

\section{Collisionless dark matter halos}

\subsection{Axially symmetric cold collapse}

Caustics are singularities in phase space (e.g. Hogan 1999). 
For example, it may be
possible to observe a radial caustic in the galaxy distribution
surrounding clusters as material on the turnaround shell is projected
onto the zero velocity surface. Physical singularities in density can
occur, for example as shells of material turnaround during the 
collapse of a dark matter halo. Spherical symmetry gives rise to shells
and a singularity at the halo center whereas axial symmetry produces
a sequence of caustic rings. 

Under certain conditions 
dynamically significant caustics are a theoretical possibility. Sikivie
has calculated the radial positions and
structure of caustic rings that occur during cold and axially symmetric
collapse. Predictions from this model are often used to calculate rates
for direct and indirect detection experiments 
therefore its assumptions should be critically examined.

Cold initial conditions are essential for the formation of narrow
caustics otherwise they will be smeared out over a radial scalelength
$\approx R_{200}\sigma_{infall}/\sigma_{halo}$.  Here $R_{200}$ is the
virialised extent of the halo and $\sigma_{infall}$ is the velocity
dispersion of infalling material.  In order to achieve a significant
density enhancement, the infalling shells must have a velocity
dispersion $< 10\%$ of the velocity dispersion of the final
halo. Unfortunately such a cold collapse is globally unstable to the
radial orbit instability (Aguilar and Merritt 1990).  This is
demonstrated in Figure 1 which shows the gravitational collapse of
$10^7$ particles with initial velocity dispersion $10\%$ of the final
velocity dispersion.  As the central region virialises
the bar instability creates a prolate mass distribution with an axial
ratio larger than 2:1. This is incompatible with observational
constraints on halo shapes.

\

\centerline{\epsfysize=6.truein \epsfbox{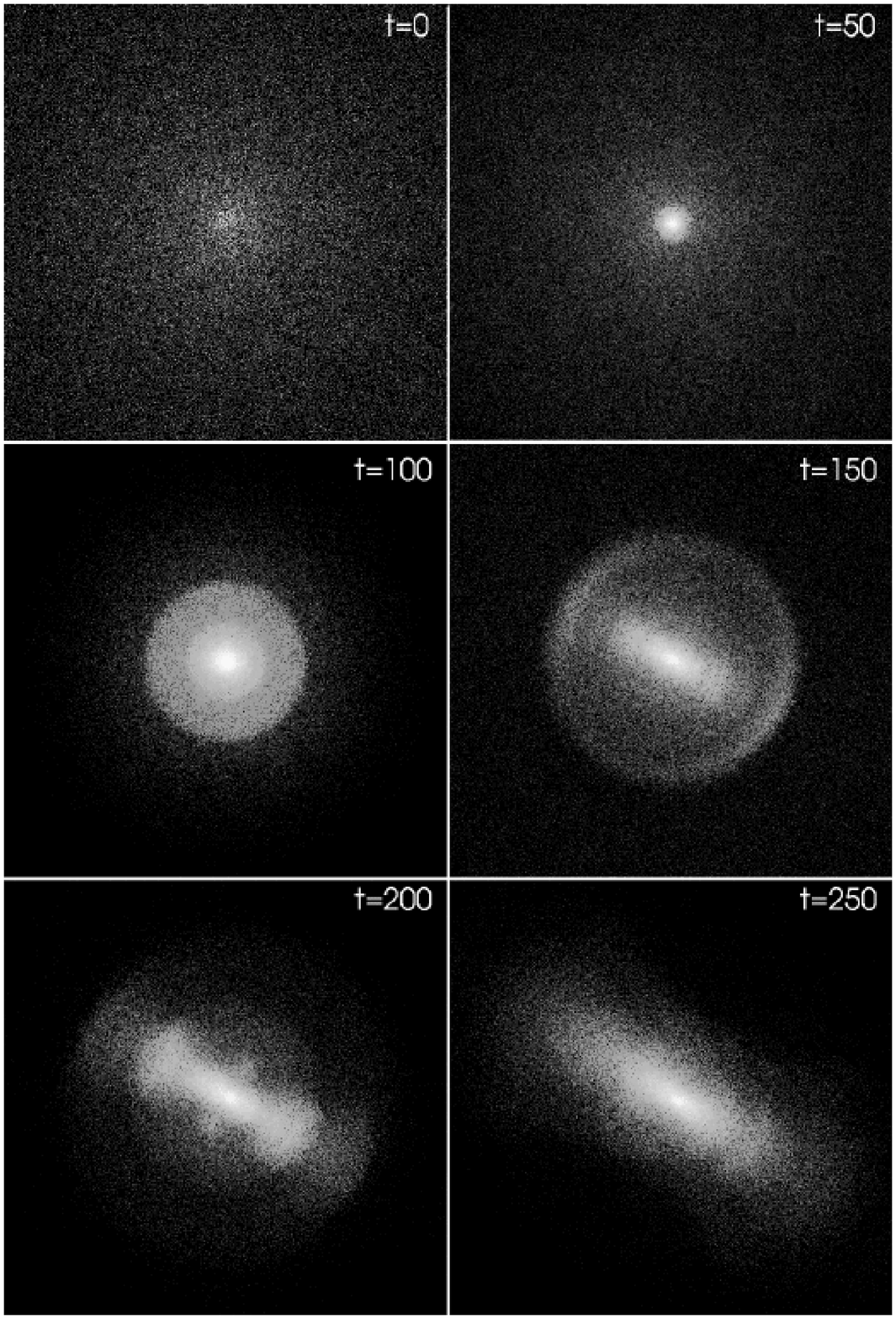}}

\

\noindent{\bf Figure 1.} \ \ The collapse of a cold, collisionless 
sphere. The final structure is a highly flattened prolate system
that results from a radial orbit instability.

\

\subsection{Hierarchical growth}

CDM halos form via a series of many mergers and accretions of dark
matter clumps along highly filamentary mass distributions. The
assumptions of symmetry and locally cold flows is an incorrect
over-simplification of the true hierarchical growth.

At a given point in a CDM halo, the ``smooth'' dark matter background
arises from material that has been tidally stripped from less massive
halos.  The infalling background of CDM is a hierarchy of
tiny halos that are unresolved in current numerical simulations. Globally, the
fraction of CDM in bound objects is expected to be 100\%, whereas
current simulations that can resolve just one level of the hierarchy 
find a value of $\sim 50\%$.

If the power spectrum was cutoff on small scales, e.g. warm or hot dark 
matter, then locally cold material would fall onto halos giving rise to
caustic structures. These features would be phase wrapped only a few times 
and it is possible that they have observable consequences.
The power spectrum of fluctuations in the CDM model allows small dense dark
matter halos to collapse at very early epochs -- the cutoff scale from the
free streaming of neutralinos is approximately $10^{-12}M_\odot$.
The first halos to collapse will have prominent caustics
although on a tiny scale compared to the solar system. Infalling material will be
scattered by various processes such as tidal fields and halo-halo encounters which smear
out the caustic sheets of CDM mini-halos (e.g. Tremaine 1999, Helmi and White 1999).

We have used numerical simulations to investigate the structure of dark matter
halos across a range of mass scales from $10^{7}M_\odot$ to
$10^{15}M_\odot$. Rather than run many simulations at low resolution we have
simulated several at very high resolution. One of the key results that we find
is that substructure within hierarchical models is essentially scale free. The
distribution and orbital properties of ``halos-within-halos'' is independent
of halo mass -- CDM galaxy halos contain literally thousands of halos more
massive than those that surround the dwarf satellite galaxies in our own halo,
Draco and Ursa-Minor.

Figure 2 shows the formation of a single CDM halo resolved with $10^6$
particles and force resolution $<0.001R_{200}$.  The substructure
visible within the CDM halo in Figure 2 is nearly self-similar.  If we
could zoom into a single subhalo it would appear like a scaled version
of the entire system.

\

\centerline{\epsfysize=5.9truein \epsfbox{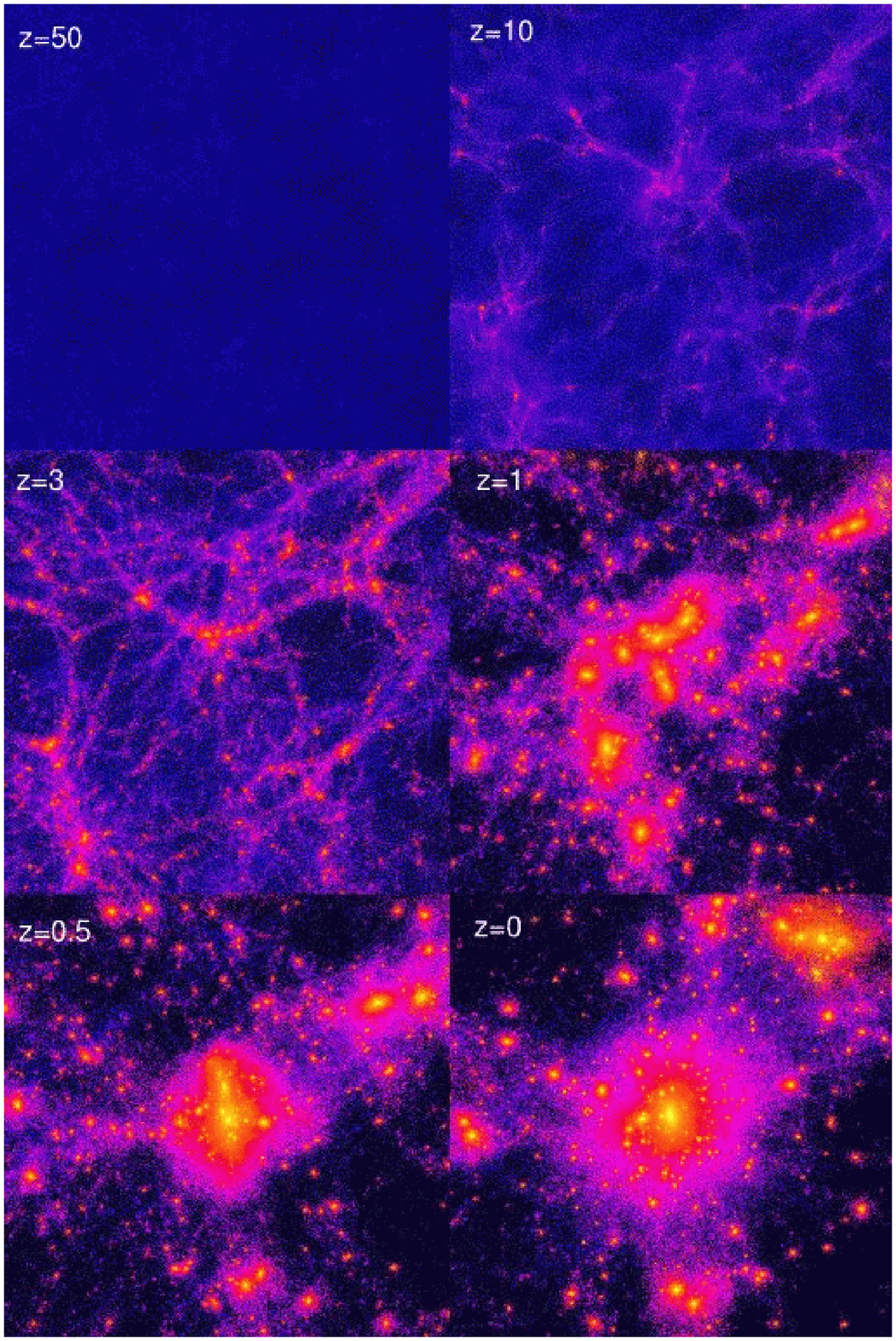}}

\

\noindent{\bf Figure 2.} \ \ The hierarchical collapse of a CDM halo. The
grey scale shows the local density of dark matter at the indicated redshifts.
The size of each panel is 4 comoving Mpc for H=50 km/s/Mpc.

\


\centerline{\epsfysize=4.3truein \epsfbox{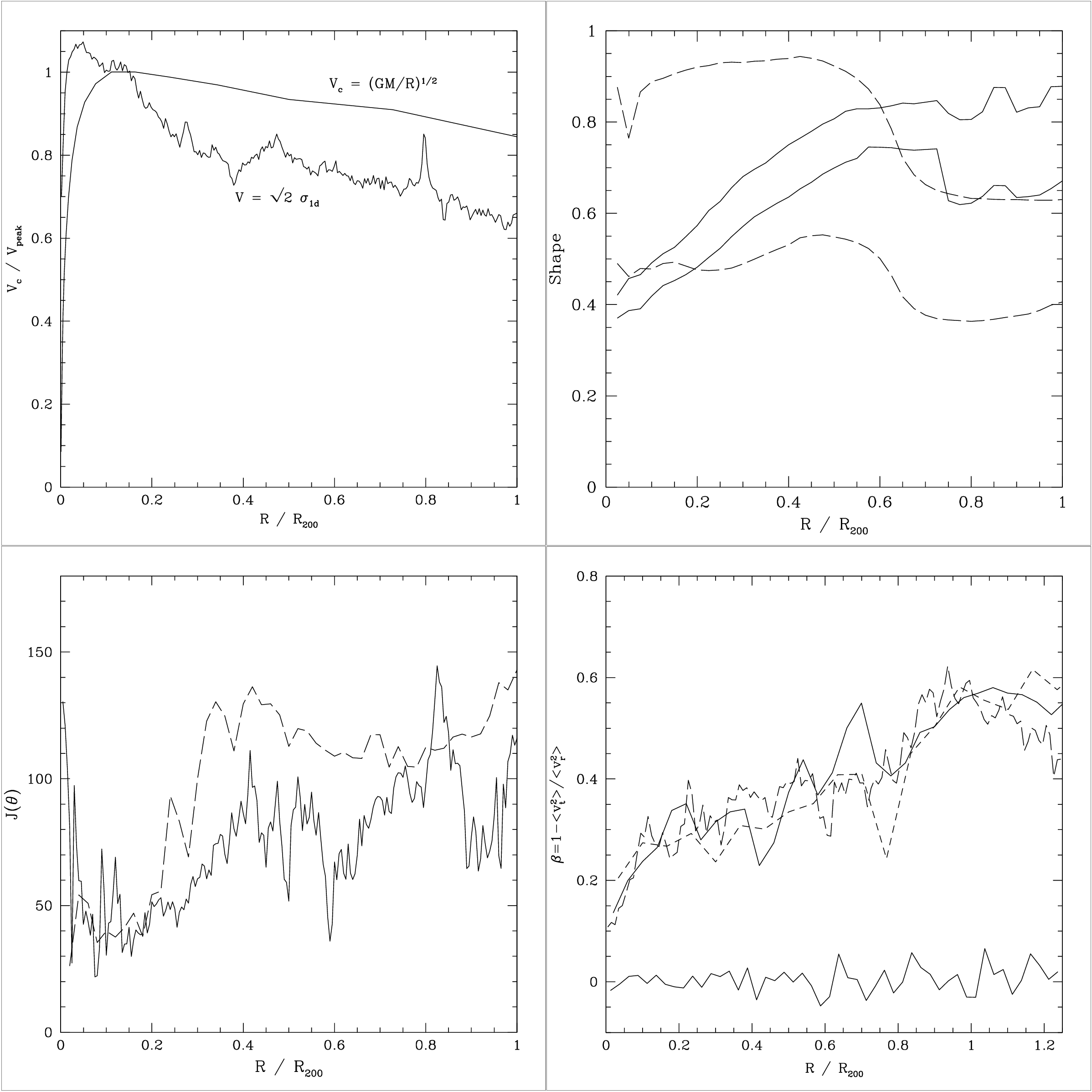}}

\noindent{\bf Figure 3.} \ \ From top left to bottom right we plot the spherically
averaged rotation curve and and 1d velocity dispersion profile of a CDM halo, 
the short to long axis ratios of two CDM halos, the angular momentum vector of
these two halos and the anisotropy parameter within a single halo
simulated at different resolutions. From these plots it is clear that
the global properties of halos are complex and axial symmetry is not preserved.

\

Finally, we show the 
structure in velocity space as a function of location within a CDM halo.
We have taken 3 cubes at different positions in the halo that
are smooth in density space. Each cube contains 2000 particles and has no
bound subhalos within them as can been seen in the central panels.  The left
panel shows the histogram of one component of the distribution of particle
speeds in each cube -- the curve shows a Maxwellian distribution with $\sigma$
set equal to the spherically averaged 
value at $0.1R_{200}$. The right panels show the $x$ and
$y$ components of velocity plotted against each other. 
In the outer halo streams of tidally stripped particles have
made just one or two complete orbits and they remain very distinct and
cold features in velocity space. 
As one moves from the outer to the inner halo, the streams
are significantly more wound up in phase space and the velocity distribution
becomes close to a Maxwellian. However this is purely a resolution effect
since higher resolution would resolve smaller halos and a network of cold
streams at the solar radius.

\

\centerline{\epsfysize=4.5truein \epsfbox{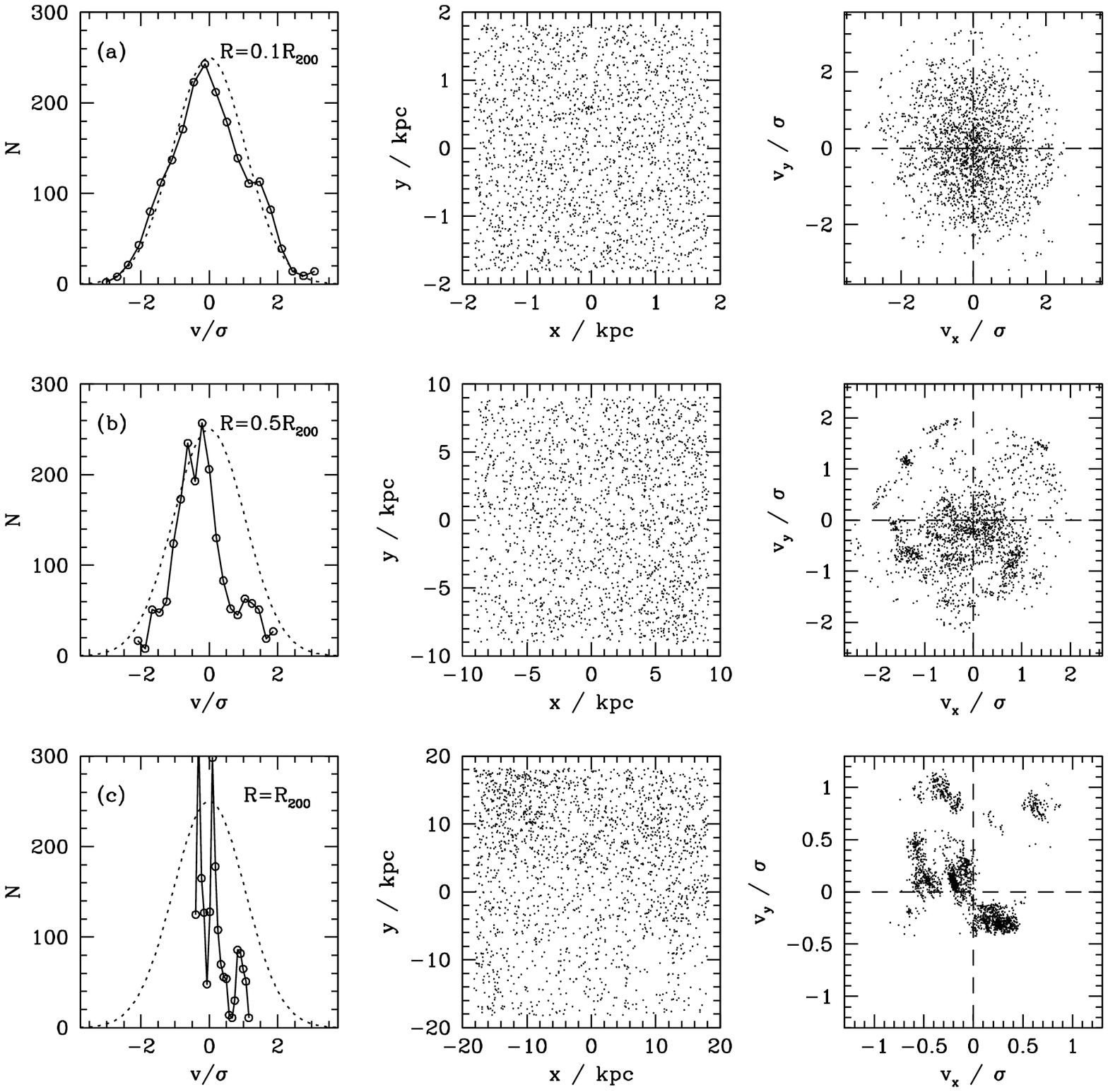}}

\noindent{\bf Figure 6.} \ \ The left panels show velocity histograms of
particles extracted from small patches of the halo where the particles
are smoothly distributed in density space (center panel) but show
complex velocity distributions (right panels).

\

\section{Summary}

We have presented preliminary results from simulations of CDM halos
that are the highest resolution possible using todays
supercomputers. These simulations have just begun to resolve
sub-kiloparsec scales, giving rise to several important predictions
on the global structure and substructure of collisionless dark matter
halos. The steep central cusps and abundant dark matter substructure
of CDM halos are fundamental and testable predictions.

The phase space structure of halos resulting from complex merger
histories are just begining to be resolved with current simulations.
We find that tidally stripped debris leaves characteristic structure
in phase space and the velocity distributions depends sensitively on
the location of the observer within triaxial CDM halos 
(Evans etal 2000, Ullio etal 2000). 
However, a great deal of work remains to determine the rate and energy
distribution of CDM particles through a detector $\sim 1m^3$ 
(Copi \& Krauss 2001).

The prominence of caustics depends crucially on the small scale power
spectrum. Detectable caustic rings require very special initial
conditions that are incompatible with the CDM paradigm -- namely
axial symmetry and locally cold velocity fields.

Future simulations and analytic work will be able to answer the key
questions: Is there a smooth component of dark matter on the scale of
the solar system?  Is the earth likely to be passing through a single
cold tidal stream or the superposition of many such streams?

\end{document}